\documentstyle[aps,bm,epsfig,amssymb]{revtex}
\newcommand{\new}{}
\newcommand{\olo}{\mbox{${\cal O} (\eps,\, \delta )$}}

\newcommand{\bef}{\begin{figure}}
\newcommand{\eef}{\end{figure}}
\newcommand{\eps}{\epsilon}
\newcommand{\leb}{\left(}
\newcommand{\rib}{\right)}
\newcommand{\bei}{\begin{itemize}}
\newcommand{\eei}{\end{itemize}}
\newcommand{\bea}{\begin{eqnarray}}
\newcommand{\boldalpha}{\bm{\alpha}}
\newcommand{\boldbeta}{\bm{\beta}}
\newcommand{\boldgamma}{\bm{\gamma}}
\newcommand{\bal}{\bm{\alpha}}

\newcommand{\bbe}{\bm{\beta}}

\newcommand{\bga}{\bm{\gamma}}

\newcommand{\eea}{\end{eqnarray}}
\newcommand{\bequ}{\begin{equation}}
\newcommand{\eequ}{\end{equation}}

\begin{document} 
\title{On the relation between coupled map lattices  \\ 
and kinetic Ising models}
\author{Frank Schm\"user\thanks{e--mail: frank@mpipks-dresden.mpg.de}}
\address{Max--Planck--Institut f\"ur Physik komplexer Systeme,
N\"othnitzer Str.\ 38, D--01187 Dresden, Germany}
\author{Wolfram Just\thanks{e--mail: wolfram@chaos.gwdg.de}}
\address{Max--Planck--Institut f\"ur 
Str\"omungsforschung, Bunsenstra\ss e 10,
D--37073 G\"ottingen, Germany}
\author{and Holger Kantz}
\address{Max--Planck--Institut f\"ur Physik komplexer Systeme, 
N\"othnitzer Str.\ 38, D--01187 Dresden, Germany}
\date{October 19, 1999}
\maketitle
\begin{abstract}

A spatially one dimensional coupled map lattice possessing the same symmetries
as the Miller Huse model is introduced. Our model is studied analytically
by means of a formal perturbation expansion which uses weak coupling and the
vicinity to a symmetry breaking bifurcation point.
In parameter space four phases with different ergodic behaviour are observed.
Although the coupling in the map lattice is diffusive, antiferromagnetic
ordering is predominant. Via coarse graining the deterministic model
is mapped to a master equation which establishes an equivalence between our
system and a kinetic Ising model. Such an approach sheds some light on the
dependence of the transient behaviour on the system size and the nature of
the phase transitions.
\end{abstract}

\section{Introduction}
Since the middle of the seventies the investigation of 
deterministic chaos has become one
of the prominent fields in science, especially physics.
A lot of knowledge has been gained since that time,
in particular for low degree of
freedom systems \cite{eckma}, and a whole machinery of tools has been
developed for the diagnostics of chaotic motion. We just mention
Lyapunov exponents and fractal dimensions as the most popular
quantities. Parallel to these developments the question has
been raised to which extent the number of degrees of freedom
enters the business. Unfortunately, much less progress has been achieved
in this direction. Only few results are available and most of them are 
bound to the investigation of model systems. Within that context
coupled map lattices (CMLs) have been introduced at the end of the 
eighties as a widely studied model class \cite{kanek,crutch}.
In such models local chaos is generated by a chaotic map which is placed at
each site of a simple lattice. Spatial aspects
are introduced by coupling these local units and special 
emphasis is on the limit
of large lattice size where the dynamics becomes high dimensional.

There is just one class of many degree of freedom systems which is fairly
well understood, namely statistical mechanics at and near thermal equilibrium.
Unfortunately, the systems studied in the field of
space time chaos are often far from equilibrium so that the tools of    
equilibrium statistical mechanics may fail. 
Nevertheless, the reduction to relevant degrees of freedom, sometimes called
coarse graining, may be equally successful in both areas.
By elimination of irrelevant degrees of freedom one maps the
microscopic deterministic equation of motion to a stochastic model
where the noise captures the irrelevant information.
Such a concept, well developed in equilibrium statistical mechanics, has
also been used in nonlinear dynamical systems;
introductions can be found on the textbook level \cite{Robinson}.
In a rigorous approach coarse graining is performed by suitable
partitions of the phase space and there are results for 
particular coupled map lattices available (cf.\ \cite{buni,brikup}).
Unfortunately, such schemes are limited to some perturbative regime
and are technically extremely difficult to apply. 
Henceforth, sometimes more physically
motivated coarse grainings are used \cite{bruessel} relaxing the amount
of rigour a little bit.

The just mentioned 
statistical methods become especially relevant in the study of phase
transitions in CMLs \cite{bohr}. Qualitative changes in the dynamical
behaviour may be related to phase transition like scenarios in the
corresponding coarse grained description. Prominent examples for such
phenomena occur in the models introduced by Sakaguchi \cite{saka} and
Miller and Huse \cite{mihu}. To keep the paper self contained and as
a motivation for the construction of our model we shortly review
basic features of the latter model. 

In order to mimic a phase transition
in a two dimensional Ising model, the chaotic antisymmetric map
depicted in figure \ref{figmamh} was placed onto a square lattice
and coupled to its four nearest neighbours
\begin{equation}
x_{i\, j}^{t+1} := \leb 1 - \eps \rib  \phi \leb  
x_{i\, j}^t \rib  +  
\frac{\eps}{4}  \sum_{k,l=\pm 1}  \phi\leb x_{i+k\, \, j+l}^t \rib \quad .
\label{eqmihuintro}
\end{equation}
Performing a coarse graining according to the sign of the phase space
variables
\bequ
\alpha_{i\, j}^t =  \left\{ \begin{array}{
l @{\, ,\quad \;
\mbox{if} \quad x_{i\, j}^t} r} 
+1 & \geq 0 \\ -1 & < 0 \end{array} \right. 
\label{eqcoarsesi}
\end{equation}
numerical simulations indicate a phase transition if the coupling strength
exceeds a critical value $\eps_{crit.} \approx 0.82 $ 
(cf.~figure \ref{figmamh}). Extensive numerical simulations 
indicate \cite{chate} that the 
phase transition is continuous. However, it is doubtful whether
the transition belongs to the Ising universality class, because the results
for the critical exponents are inconclusive. In particular, their values
depend on whether the CML is updated synchronously or
asynchronously. One can summarise that the phase transition of the Miller Huse
model is still far from being understood, in particular since
no quantitative description of the spin dynamics could be derived. 
In order to reach some progress in this direction we here
introduce and investigate a slightly different model system with
analytical methods.
\bef
\begin{center}
\mbox{\epsfig{figure=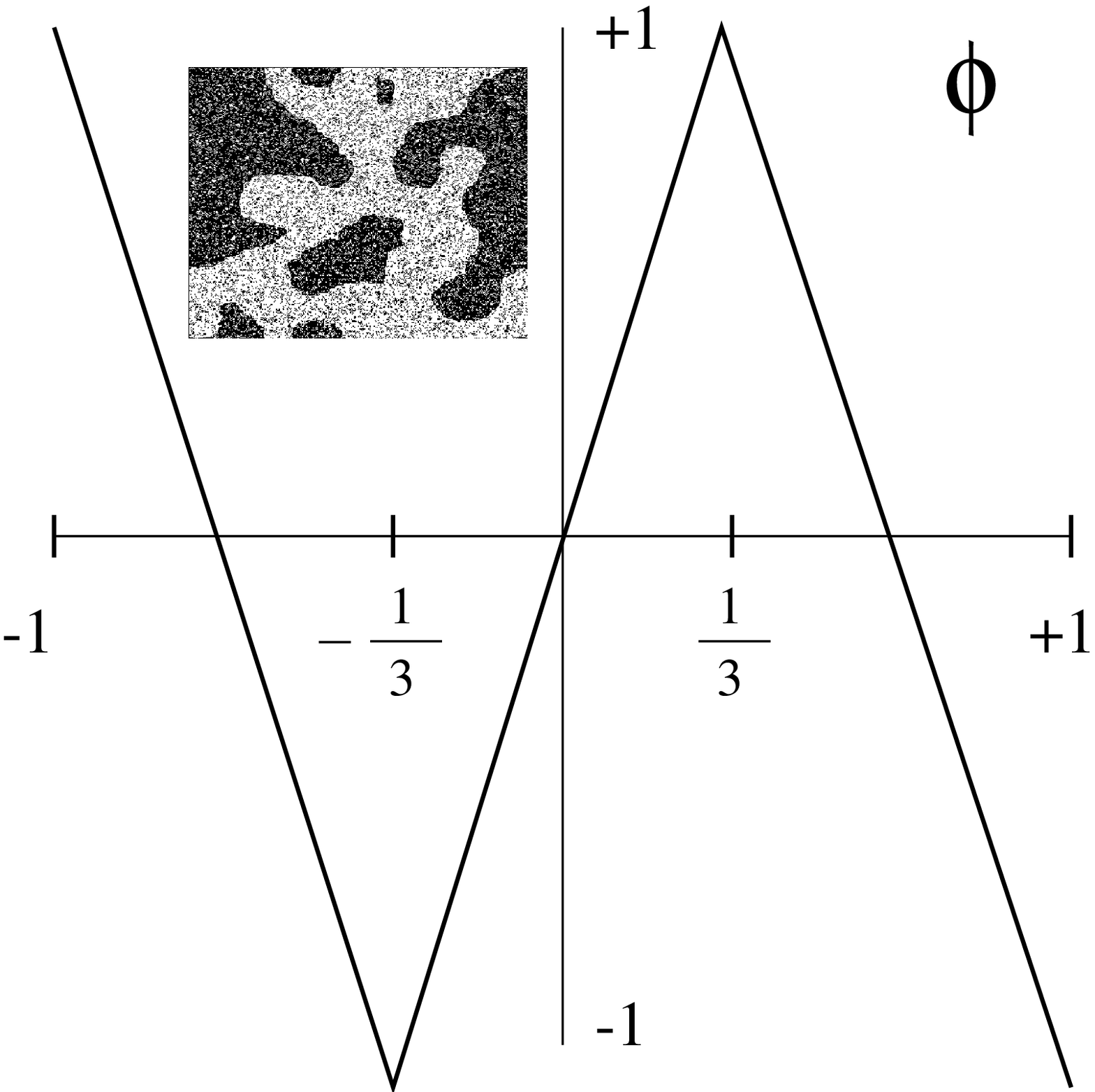,width=50mm}}
\end{center} 
\caption[ ]{The single site map $\phi$ of the Miller Huse model.
The inset shows a typical pattern for $\eps>\eps_{crit.}$
(white/black $\alpha_{i\, j}=\pm 1$). \label{figmamh}}
\eef

Section \ref{secmodel} introduces our model 
as well as the setup of the perturbation expansion. For the latter
purpose transitions between sets of a suitable partition are defined.
These transitions are studied in detail in section \ref{secttrap}.
Herewith, the bifurcation diagram of our model will be developed 
in section \ref{secbifu} and
analytical expressions for the bifurcation lines are 
calculated in perturbation theory. Section \ref{seccoa} is devoted
to a systematic coarse graining of the dynamics on the basis of the
just mentioned partition. On that level the dynamics is described
in terms of a master equation which corresponds to a particular class
of kinetic Ising models. It constitutes the basis
for the investigation of the transient behaviour in section \ref{sectrans}. 
Finally, the main results of this work are
summarised. The appendices are concerned 
with parts of the perturbation expansion, 
but more details can be found in \cite{diss}.
 
\section{The model} \label{secmodel}
Let us first consider the single site map. It consists of 
a deformed antisymmetric tent map $f_{\delta}$, which is
linear on three subintervals of $[-1, \, 1]$
\begin{equation}
f_{{\delta}}(x) := \left\{ \begin{array}{r @{\; , \; \;{\rm if} 
\quad \; x  \in\,} l}
-2 - x / a  & [-1, -a  ] \\ x / a  \quad & (-a  , a  ) \\ 2 - x / a  & 
[a  ,1] \end{array}
\right. \; , \quad  a   :=  \frac{1}{2 - { \delta}} \quad .
\label{eqdeffd}
\end{equation}
Because of $f_{{\delta}}(1) = \delta$ the parameter $\delta$ 
determines whether transitions between the intervals
$[-1,0]=:J(-1)$ and $[0,1]=:J(+1)$ are possible.
Note that the Miller Huse map is obtained as a special case,
$\phi = f_{\delta = -1}$. The introduction of $a$ in eq.~(\ref{eqdeffd})
ensures that the modulus of the derivative
of $f_{{\delta}}$ is constant on the whole interval. 
Figure \ref{figourmap} shows the function
$f_{\delta}$ for small positive and negative $\delta$.
For $\delta \geq 0$ the single site map has
two coexisting attractors, the intervals 
$[-1, \, \delta]$ and $[\delta, \, 1]$,
whereas for  $\delta < 0$ only one attractor, the
interval $[-1, \, 1]$, is present. 
\bef
\begin{center}
\mbox{\epsfig{figure=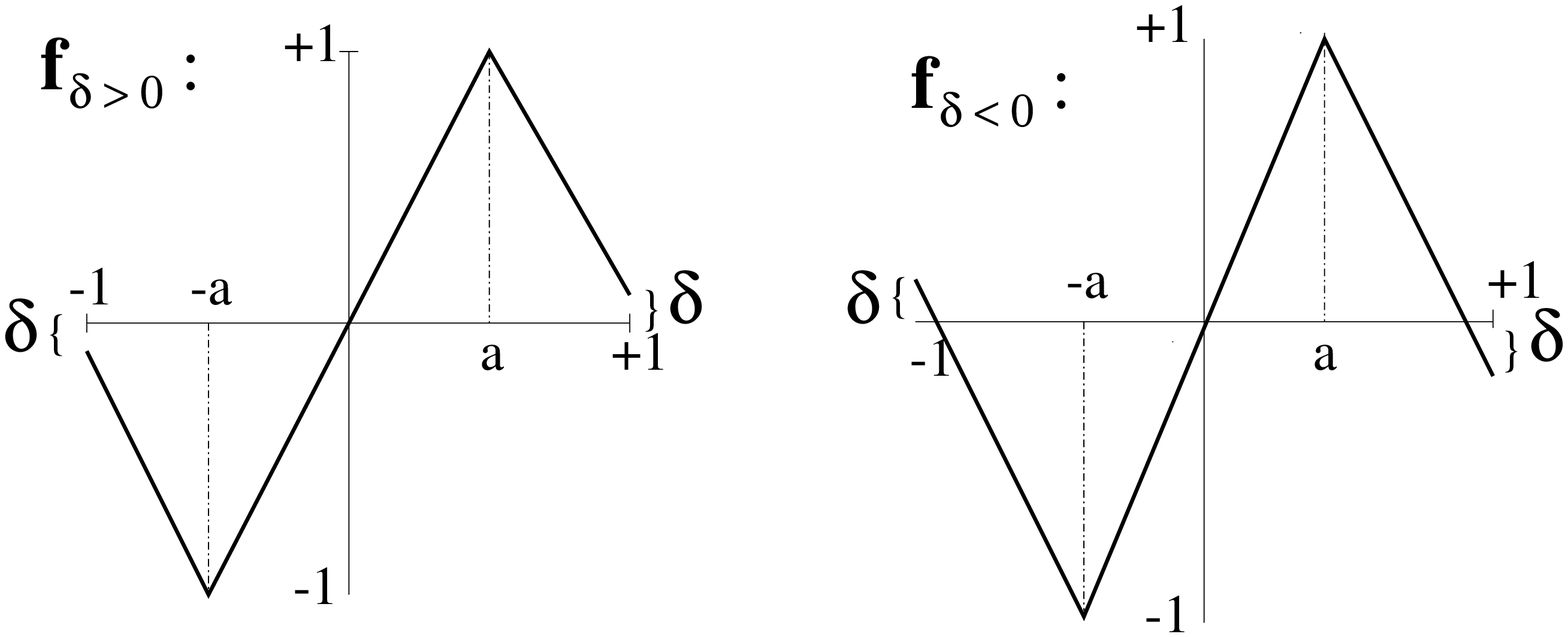,width=100mm}}
\end{center}
\caption{The deformed antisymmetric tent map $f_{\delta}$.}
\label{figourmap}
\eef

The CML which is studied in this article
is defined on a one dimensional lattice (chain) of length $N$.
Nearest neighbours are coupled in a standard
''diffusive'' way with periodic boundary conditions
\bea 
{\mathbf{T}}_{{\eps}, { \delta}}& : & [-1, +1]^N \rightarrow  [-1, +1]^N \; ,
\nonumber \\ 
\left[ \mathbf{T_{{\eps}, { \delta}}({\mathbf{x}})} \right]_{i} & := & (1 - 
{\eps})\, f_{{ \delta}}(x_i) +
\frac{\eps}{2} \, (f_{{ \delta}}(x_{i-1}) + f_{{ \delta}}(x_{i+1}))
\label{eqcml} \quad .
\eea
The parameter $\eps$ denotes the coupling strength. Because of the
single site map and the diffusive coupling the CML 
${\mathbf{T}}_{{ \eps },{\delta}}$ has the symmetry
${\mathbf{T}}_{{ \eps },{\delta}}(- {\mathbf{x}}) =  - {\mathbf{T}}_{{ \eps },
{\delta}}({\mathbf{x}})$.
Furthermore, translation invariance on the one dimensional
lattice holds, because periodic boundary conditions have been imposed.

Since we are going to
perform a perturbation theory with $\eps, \, | \delta | \ll 1$, we
first consider the CML with $\eps = \delta = 0$. In this 
case the model can be solved trivially.
The non--deformed antisymmetric tent map $f_0$ has the
two attractors $J(-1)=[-1, \, 0]$ and 
$J(+1)=[0, \, 1]$. Therefore, $N$ uncoupled maps
$f_0$ have $2^N$ coexisting attractors, each one an $N$ dimensional cube of
edge length one
\begin{equation}
I_{\boldalpha} :=   J({\alpha}_1) \times J({\alpha}_2) \times \dots  \times
J({\alpha}_N) \; \quad. \label{kub1}
\end{equation}
We distinguish these cubes $I_{\bal}$ by an 
$N$ dimensional index vector $\bal =  (\alpha_1, \alpha_2, \dots \alpha_N)$
where $\alpha_i \in \{-1, \, +1 \}$.
The natural measure on each cube is the Lebesgue measure.
As we will see these cubes become important building blocks of the perturbation
theory and the starting point of a coarse grained description of the CML
${\mathbf{T}}_{{\eps}, { \delta}}$. 

From a dynamical systems point 
of view we are mainly interested 
in ergodic properties of the CML, i.~e.~the
number of coexisting attractors 
and their location for given small parameters $\eps, \,
\delta$. An important observation is that in the perturbative regime a typical
orbit stays for many iterations within a cube $I_{\bal}$, before it
possibly enters another cube $I_{\bbe}$. Therefore, in perturbation theory any
attractor of the CML ${\mathbf{T}}_{{\eps}, { \delta}}$ is a union of cubes
$I_{\bal}$, if one neglects sets with volume $\olo$. Hence, the dynamics
is sufficiently characterised by transitions $ I_{\bal} \rightarrow
I_{\boldbeta}$ between cubes. 

Of course we have to be more definite with what we mean by a transition.
In order that a phase space point can be mapped from a cube $I_{\bal}$
to a cube $I_{\bbe}$ ($\bal \neq \bbe$) the image of the former has
to intersect the latter. Hence the {\em overlap set}
\begin{equation} 
OV_{\bal, \, \bbe} := {\mathbf{T}}_{{ \eps },{\delta}}( I_{\bal} ) \cap
I_{\bbe} 
\label{overlapset}
\end{equation}
plays an important role. A necessary condition for a point to
migrate from $I_{\bal}$ to $I_{\bbe}$ is a non--empty
overlap set $OV_{\bal, \, \bbe}$.
Since in perturbation theory the set ${\mathbf{T}}_{{ \eps
},{\delta}}(I_{\boldalpha} )$ is a weakly deformed cube $I_{\bal}$, 
the set $OV_{\bal, \, \bbe}$ can at most have a volume of size $\olo$.
However, the condition on the overlap set is far from being sufficient
because one has to ensure that typical orbits can reach this set upon
their itinerary. For that purpose two additional conditions have
to be imposed.

First, we have to ensure that points from the inner 
part\footnote{For our perturbative treatment we define the inner part
as the set of all ${\mathbf{x}} \in I_{\bal}$ which have at least a small
fixed positive distance $d$ from the boundary, where the quantity $d$
does not depend on the expansion parameters $\eps$ and $\delta$.} of
$I_{\bal}$ reach the overlap set. For that reason we consider
the pre--images of $OV_{\bal, \, \bbe}$ of 
various generation that are contained in $I_{\bal}$ 
\bea
{\mathbf{T}}_{{ \eps }, {\delta}}^{-1}(OV_{\bal , \, \bbe}) & := &
\{ {\mathbf{x}} \in I_{\bal } \; | \; {\mathbf{T}}_{{ \eps }, {\delta}}(
{\mathbf{x}}) \in OV_{\bal , \, \bbe} \}  \nonumber \\
{\mathbf{T}}_{{ \eps }, {\delta}}^{-k}(OV_{\bal, \, \bbe}) & := &
\{ {\mathbf{x}} \in I_{\bal} \; | \; {\mathbf{T}}_{{ \eps }, {\delta}}(
{\mathbf{x}}) \in {\mathbf{T}}_{{ \eps }, {\delta}}^{-(k-1)}(OV_{\bal, \, 
\bbe})  \}
\, , \hspace{.6cm} k=2,3,\dots \quad .
\label{eqdefurb1}
\eea
For some finite $k$ the pre--image set ${\mathbf{T}}_{{ \eps },
{\delta}}^{-k}(OV_{\bal, \, \bbe})$ should intersect the inner part of
$I_{\bal}$, so that points from the inner part of $I_{\bal}$ can reach the
overlap set $ OV_{\bal, \, \bbe} $.\footnote{Since for $\eps =\delta =0$ 
the natural measure on each cube is the Lebesgue measure, 
in the perturbative regime the map ${\mathbf{T}}_{{ \eps },{\delta}}$
distributes the points of an orbit rather uniformly within a cube $I_{\bal}$.
Therefore, in determining the orbit dynamics it suffices to use topological 
methods like the calculation of pre--image sets.}

The points of the set $OV_{\bal, \, \bbe}$ are near the surface of the 
cube $I_{\bbe}$ within a distance of order $\olo$. 
The second condition demands that points 
from a subset of  $ OV_{\bal, \, \bbe} $ with finite 
Lebesgue measure reach the inner part of 
the cube $I_{\bbe}$ directly under further iteration. 
The two conditions for a transition  $ I_{\bal} \rightarrow I_{\bbe}$ 
ensure that the transition is
possible for a set of finite Lebesgue measure that 
is located in the inner part of $ I_{\bal}$.

\section{Transitions in perturbation theory} \label{secttrap}
In what follows we consider the CML ${\mathbf{T}}_{{ \eps },{\delta}}$ for
arbitrary but fixed
lattice size $N$. We would like to know which transitions 
$ I_{\bal} \rightarrow I_{\bbe}$ are possible for given parameters $\eps$,
$\delta$. In the spirit of perturbation theory we confine ourselves to
dominant transitions. Those are transitions where the cubes $I_{\bal}$ 
and $I_{\bbe}$ share an $(N-1)$ dimensional surface. Then, the
volume of the overlap set $ OV_{\bal, \, \bbe} $ can be greater by 
a factor $1 / \eps $ or $1 / | \delta |$ in comparison to the case 
without a common surface. Consequently, the $N$ dimensional
index vectors $\bal$ and $\bbe$ only differ in one component, the transition
index $\alpha_i$. In such a transition $ I_{\bal} \rightarrow I_{\bbe}$ 
the $x_i$ coordinate of the phase 
space orbit $\{ {\mathbf{x}}^t\}$ 
changes its sign. {\new Transitions of higher order in which two or 
more coordinates simultaneously change their sign will not be considered in
this article, because their rates are smaller by a factor of the order $\olo$
in comparison to the dominant transitions.} 

In perturbation theory, for a dominant transition only the neighbouring
indices of the transition index, $\alpha_{i-1}$ and $\alpha_{i+1}$, are
relevant, because of the nearest neighbour interaction
of the map ${\mathbf{T}}_{{ \eps }, {\delta}}$ (cf.\ eq.~(\ref{eqcml})). 
In addition, the influence of the two neighbouring coordinates $x_{i-1}$ and 
$x_{i+1}$ on the $x_i$ coordinate 
is predominant for a finite number of iterations, since
interactions with lattice sites farther away are suppressed by the small 
coupling strength $\eps $. More precisely, within first order perturbation
theory the overlap sets $ OV_{\bal, \, \bbe}$ and their pre--image sets 
can be approximated by the following product sets
(cf.~appendix \ref{appreduct})
\bea
OV_{\bal\, , \bbe} & = & OV^{(3)}_{\alpha_{i-1} \, \alpha_i \,
\alpha_{i+1} , \beta_{i-1} \, 
\beta_i \, \beta_{i+1}} \times I^{(N-3)}_{\alpha_1 \,
\alpha_2 \dots \alpha_{i-2} \, \alpha_{i+2} 
\dots \alpha_{N}} \; , \nonumber \\
{\mathbf{T}}_{\eps , \delta }^{-k} (OV_{\bal, \, \bbe}) & = &
\left[ {\mathbf{T}}_{\eps , \delta }^{(3)} \right]^{-k} 
\leb OV^{(3)}_{\alpha_{i-1} \, \alpha_i \,
\alpha_{i+1}, \beta_{i-1} \, \beta_i \,
\beta_{i+1} } \rib \; \times  \; I^{(N-3)}_{\alpha_1 \,
\alpha_2 \dots \alpha_{i-2} \, \alpha_{i+2} \dots \alpha_{N}} \, ,
 \quad \quad  k \geq 1 \;  \quad .
\label{eqprourbn4}
\eea
Here $OV^{(3)}_{\alpha_{i-1} \, \alpha_i \, \alpha_{i+1} ,
\beta_{i-1} \, \beta_i \, \beta_{i+1}} $ denotes 
a three dimensional projection of
the full overlap set which contains the coordinates $x_{i-1}$, $x_i$ and
$x_{i+1}$, and ${\mathbf{T}}_{\eps , \delta }^{(3)}$ 
denotes the map lattice for $N=3$. 
The $(N-3)$ remaining coordinates are contained in the 
$(N-3)$ dimensional cube $I^{(N-3)}_{\alpha_1 \,\alpha_2 \dots \alpha_{i-2} 
\, \alpha_{i+2} \dots \alpha_{N}}$. Effectively, we have herewith reduced the
transition in a map lattice of size $N$ to a transition in a 
map lattice of size three,
because $(N-3)$ coordinates play only a spectator role. 
Put differently, the CML ${\mathbf{T}}_{{ \eps },{\delta}}$ reaches  
already its full complexity for $N=3$, if one stays in the perturbative regime.

For symmetry reasons one can identify three different
types of transitions $ I_{\bal} \rightarrow I_{\bbe}$:
\begin{description}
\item[Type (a):] the three indices $\alpha_{i-1}$, $\alpha_i$ and
$\alpha_{i+1}$ are equal, e.~g.
\[ 
I_{\dots , \, +1, \, +1, \, +1,\dots} \rightarrow I_{\dots , \, +1, \, -1, \, 
+1, \dots} \quad .
\]
\item[Type (b):] the two neighbouring indices $\alpha_{i-1}$ and 
$\alpha_{i+1}$ are different from each other, e.~g.~
\[ 
I_{\dots , \, -1, \, +1, \, +1,\dots} \rightarrow I_{\dots , \, -1, \, -1, \, 
+1, \dots} \quad .
\]
\item[Type (c):] the neighbouring indices $\alpha_{i-1}$ and $\alpha_{i+1}$
differ from the transition index $\alpha_i$, e.~g.~
\[ 
I_{\dots , \, +1, \, -1, \, +1,\dots} \rightarrow I_{\dots , \, +1, \, +1, \, 
+1, \dots} \quad .
\]
\end{description}
Transitions of type (c) are inverse to those of type (a).

Because of the conditions mentioned in the last section transitions 
are possible only if the deformation is small enough, $\delta <
\delta_{crit.} \leb \eps \rib$. Within perturbation theory we obtain for the
different critical values
\bea
{\rm type \; (a) \,:}\; \delta_a = 0 \; , 
\quad {\rm type \; (b) \,:}\; \delta_b
= - \frac{2 \, \eps}{3} \; ,  \quad {\rm type \; (c) \,:}\; \delta_c
= - \frac{4 \, \eps}{3} \;  \quad .
\label{eqcrival}
\eea
{\new One might wonder why transitions (b) and (c) do not appear for negative
$\delta$ above the critical value. The main reason is that
despite of the existence of a non--empty overlap set trajectories do not 
reach this overlap since there exists a forbidden region in phase space
called the ''blind volume''. Points belonging to the blind volume have no
pre--images themselves. The blind volume is non--empty, since the map 
${\mathbf{T}}_{\eps , \delta }$ is not surjective for finite coupling $\eps$.
The actual} calculation of critical $\delta $ 
values necessitates rather involved 
geometric constructions in phase
space, since one must determine the location of the pre--image sets    
${\mathbf{T}}_{\eps , \delta }^{-k} (OV_{\bal, \, \bbe})$ in $I_{\bal}$. 
Hence, details are postponed to appendix \ref{appdeltac}. 
The smaller the deformation parameter $\delta$, the
more transitions become possible as
can be guessed from the geometry of the map (cf.\ figure \ref{figourmap}).
On the other hand, increasing coupling constant $\eps $ inhibits
transitions, since eventually only the transition of
type (a) remains feasible for fixed negative $\delta $. 
Such an observation contradicts somehow the
intuitive reasoning about a ''coupling'' of lattice sites. The inhibition
effect for transitions is caused by the existence of a  
''blind volume'' in the cube $I_{\bal}$ that grows with 
$\eps$ (cf.~appendix \ref{appdeltac}). 

{\new At this stage some remarks on the accuracy of our perturbative
approach seem to be in order. Since we neglect transitions of higher
order our arguments are not rigorous. In fact for a real proof the
complete absence of such transitions must be shown. For the case of
two coupled maps, $N=2$, such a step can be easily supplemented
(cf.\ appendix \ref{apphighord}) and we infer that one might be able
to perform similar but more involved computations in higher dimensional 
cases too. Nevertheless, even if these transitions are mathematically possible
their effect may be small e.g.\ taking a time scale argument into account.}
\section{The bifurcation scenario} \label{secbifu}

Eq.~(\ref{eqcrival}) determines four regions in the 
$\leb \eps, \, \delta \rib $ parameter plane where different transitions
are possible (cf.\ figure \ref{figbifu3}). Crossing these lines a bifurcation 
occurs. Between different regions the number of coexisting attractors
and their location change. {\new Determining attractors in the strict 
mathematical sense just from the knowledge of the dominant transitions 
faces however some problems.
For, neglecting sets with volume $\olo$ a union of cubes
$A$, which an orbit can not leave through a dominant transition, is a
candidate for an attractor. But, possibly an orbit can escape from this set
through a transition of higher order in perturbation theory. 
Then the set $A$ would not be an attractor in a strict mathematical 
sense. However, one can put forward the following time 
scale argument: in perturbation theory transitions 
through which an orbit can leave $A$ occur on a rather large time 
scale in comparison to the relatively fast dominant transitions through 
which the orbit is pulled back to the set $A$ again. Because of this
intermittent dynamics the set $A$ is a core region of a possibly bigger
attractor, i.~e.~ the sets $A$ are the carriers of most of the 
natural measure of these attractors. For brevity we will call these sets $A$
''attractors'' in the following.} 

If we neglect sets with volume
$\olo$ we can identify an attractor $A$ with a union of cubes $I_{\bal}$.
\bef
\begin{center}
\mbox{\epsfig{figure=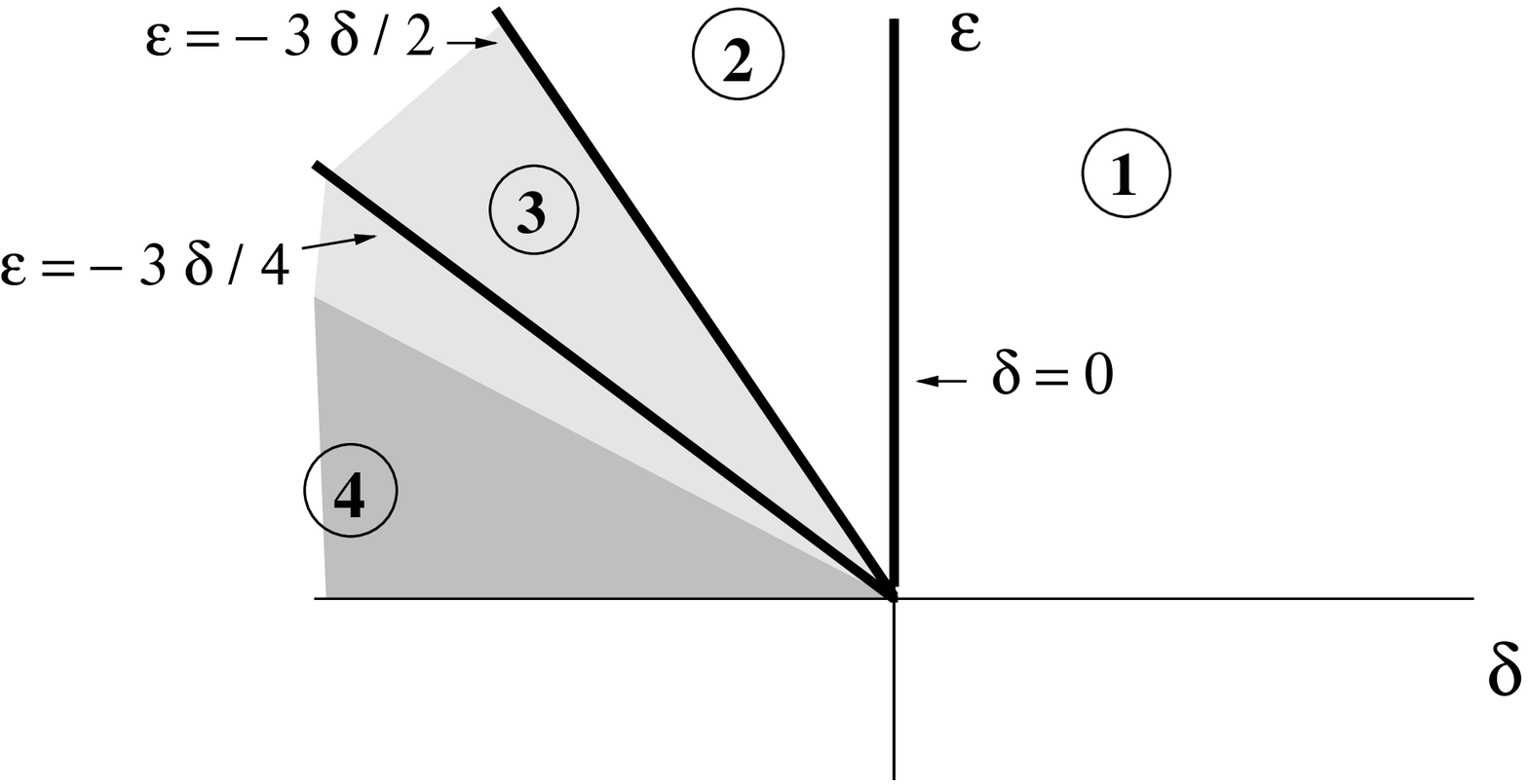,width=80mm}}
\end{center}
\caption[ ]{Diagrammatic view of the bifurcation diagram for
the CML $\mathbf{T}_{{ \eps }, {\delta}}$ according to eq.~(\ref{eqcrival}). 
The numbers in the four parameter regions refer to the text. 
Gray--shading indicates the type of coupling in the corresponding
kinetic Ising model, antiferromagnetic (light) resp.~ferromagnetic (dark)
(cf.\ section \ref{seccoa}). \label{figbifu3}} 
\eef
\begin{description}
\item[Region 1 ($\delta \geq 0$):]
No transition $ I_{\bal} \rightarrow I_{\bbe}$ is possible. Therefore, each
cube $I_{\bal}$ is an attractor so that there are $2^N$ coexisting attractors. 
\item[Region 2 ($-2 \, \eps/ 3 \leq \delta < 0$):]
Only transition type (a) is allowed. Hence, cubes $I_{\bal}$ are 
attractors such that $\bal $ does not contain three successive ''$+1$'' or 
''$-1$''. With a combinatorial argumentation one can show for that
for long chains ($N \gg 1$) the number of coexisting attractors
increases like $((1+\sqrt{5})/2)^N$.
%
\item[Region 3 ($-4 \, \eps/ 3 \leq \delta < -2 \, 
\eps/ 3 $):]
To determine the attractors in this region, it seems necessary to anticipate
the coarse graining of the CML ${\mathbf{T}}_{\eps , \delta}$ which will be
discussed systematically in section 
\ref{seccoa}. Analogous to eq.~(\ref{eqcoarsesi})
we can view the index vector $\bal$
of a cube $I_{\bal}$ as a spin chain of length $N$, where $+1$ and $-1$ are the
possible spin states on each lattice site. 
In this way the three transition types (a), (b) and (c)
translate into three different kinds of spin flips. 
For each spin chain one can define defects in the same way as in the 
antiferromagnetic Ising model. A defect (''$1$'') occurs, if two
neighbouring spins are aligned, and no defect is present, 
if the spins point in opposite directions. Then, the just mentioned
spin flips translate into a dynamics of defects.
\begin{description}
\item[Type (a):] two adjacent defects annihilate each other, e.~g.~
\bea
{\rm spin \; chain}\; \bal:  & \makebox[8cm][l]{$\; \dots  +1, \; +1, \;  +1,
\ldots  \ldots \rightarrow \dots  +1, \; -1, \;  +1, \dots  $} & \nonumber \\
{\rm defects \; in }\; \bal: & \makebox[8cm][l]{$ \;  \; \dots 
\dots 1, \; \; \; \; 1, \, \dots \dots\, \; \;  \rightarrow \; \; \dots \dots
\; 0 , \; \; \; \;\; 0, \, \dots \dots $} & \nonumber 
\eea
\item[Type (b):] one defect diffuses to a neighbouring lattice site, e.~g.~
\bea
{\rm spin \; chain}\; \bal:  & \makebox[8cm][l]{$\; \dots  +1, \; +1, \;  -1,
\ldots  \ldots \rightarrow \dots  +1, \; -1, \;  -1, \dots  $} & \nonumber \\
{\rm defects \; in }\; \bal: & \makebox[8cm][l]{$ \;  \; \dots 
\dots 1, \; \; \; \; 0, \, \dots \dots\, \; \;  \rightarrow \; \; \dots \dots
\; 0 , \; \; \; \;\; 1, \, \dots \dots $} & \nonumber 
\eea
\item[Type (c):] two adjacent defects are generated simultaneously, e.~g.~
\bea
{\rm spin \; chain}\; \bal:  & \makebox[8cm][l]{$\; \dots  +1, \; -1, \;  +1,
\ldots  \ldots \rightarrow \dots  +1, \; +1, \;  +1, \dots  $} & \nonumber \\
{\rm defects \; in }\; \bal: & \makebox[8cm][l]{$ \;  \; \dots 
\dots 0, \; \; \; \; 0, \, \dots \dots\, \; \;  \rightarrow \; \; \dots \dots
\; 1 , \; \; \; \;\; 1, \, \dots \dots $} & \nonumber 
\eea
\end{description}  

For the determination of the attractors in the present parameter region we
consider an orbit $\{ {\mathbf{x}}^t \}$ of the CML which performs successive
transitions $I_{\bal} \rightarrow I_{\bbe}$.  Each transition
changes the corresponding spin chain $\bal $ and its defects.
Since transitions of type (c) are forbidden,
defects can diffuse and annihilate in pairs only, and
the number of defects decreases monotonically.

If the size of the system $N$ is even the chain contains an even number 
of defects. An orbit $\{ {\mathbf{x}}^t \}$ 
migrates between cubes $I_{\bal}$, until all
defects have annihilated each other. Then, the orbit can not execute any
further transition of type (a) or (b). Therefore, there are two attractors, the
cubes $I_{(\,  { +1 },  {-1},  { +1 },   {-1}, \dots ,  {-1}, { +1 },  
{-1} \,)} $ and $ I_{(\,   {-1}, { +1 },   {-1},  { +1 }, \dots , { 
+1 },  {-1}, { +1 } \,)}$. {\new 
Extensive numerical simulations indicate that for $N$ even
each cube $I_{(\,  { +1 },  {-1},  { +1 },   {-1}, \dots ,  {-1}, {
+1  }, {-1} \,)} $ and $ I_{(\,   {-1}, { +1 },   {-1},  { +1 }, \dots , { 
 +1 },  {-1}, { +1 } \,)}$ constitutes an attractor in the strict sense,
i.~e.~for $\delta > - 4 \,\eps / 3$ no 
additional transition of higher order perturbation theory 
is present (cf.~also appendix \ref{apphighord})} 

For $N$ odd the number of defects in $\bal$ is odd. Consequently,
at the end of the transient dynamics one defect remains.
Since the defect can change its location
via a transition of type (b), the attractor is the union of all 
$2 \, N$ cubes $I_{\bal}$ for which $\bal$ contains a single ''$+1 \; +1$'' 
or ''$-1 \; -1$'' sequence. 

Since in both cases the ratio of the volume of the attractor to the volume of
its basin of attraction becomes very small for $N \gg 1$
one expects long transients to occur.
Section \ref{sectrans} is devoted to a more detailed study of the 
transient dynamics. Our argumentation has used the assumption
that different transitions are not correlated. 
We will come back to this problem in the next section.
\item[Region 4 ($ \delta < -4 \, \eps/ 3 $):]
All three transition types are possible. Therefore 
an orbit $\{ {\mathbf{x}}^t \}$ can visit every cube $I_{\bal}$, so that
there emerges one attractor which encompasses all cubes.   
\end{description}  
\section{Coarse graining of the CML} \label{seccoa}
Coarse graining the CML ${\mathbf{T}}_{\eps , \delta}$ one passes from orbits 
$\{ {\mathbf{x}}^t, \; t=0, \, 1, \, 
2 \dots \}$ in phase space to symbol or spin chains $\{ 
\bal^t, \; t=0, \, 1, \, 2 \dots \}$. The spin chain $\bal^t$ just
indicates the cube which contains the phase space point 
${\mathbf{x}}^t$ at time $t$ (cf.~eq.~(\ref{eqcoarsesi})\,).
If an orbit of the CML performs a transition $I_{\bal}
\rightarrow I_{\bbe}$, the state of the spin chain changes from $\bal $ to
$\bbe $. Since in perturbation theory an orbit typically circulates for 
many iterations within a cube $I_{\bal}$, the sequence 
$\{ \bal^t, \; t=0, \,
1, \, 2 \dots \}$ has a constant value for long time 
before a spin flip occurs. Altogether,
the CML is described by a {\em stochastic} spin
dynamics. First we argue that the spin dynamics is Markovian for the 
following reasons:
\bei
\item Two successive transitions $I_{\bal} \rightarrow I_{\bbe}$ and 
$I_{\bbe} \rightarrow I_{\bga}$ are uncorrelated. In the perturbative 
regime an orbit performs a highly chaotic motion within the cube $I_{\bbe}$
for many iterations before the transition to the cube $I_{\bga}$
occurs. Therefore, the memory of the preceding transition $I_{\bal} \rightarrow
I_{\bbe}$ is lost.
\item A transition $I_{\bal} \rightarrow I_{\bbe}$ is equally
probable for each iteration step. For a transition
the orbit point $ {\mathbf{x}}^t $ must hit a characteristic
set in the inner part of the cube $I_{\bal}$ which consists of pre--images of
the overlap $OV_{\bal, \, \bbe}$. During its stay within the
cube $I_{\bal}$ the orbit is distributed uniformly within $I_{\bal}$, since
for $\eps = \delta = 0$ the natural measure on $I_{\bal}$ is the Lebesgue
measure. Consequently, the
probability for the orbit to hit the characteristic set is
independent of time. 
\eei
Since the spin dynamics is Markovian at least approximately, 
the probability $p_{\bal}(t)$ that the spin chain is in state
$\bal$ at time $t$ obeys a master
equation with transition probabilities $  w \leb \bbe \, | \, \bal \rib  $ for
a spin flip $\bal \rightarrow \bbe $
\bequ
p_{\bal}(t+1) = p_{\bal}(t) + \sum_{\bbe \neq \bal} \left[ w ( \bal \, | \,
\bbe) \cdot  p_{\bbe}(t) - w ( \bbe \, | \, \bal) \cdot  p_{\bal}(t) \right] 
\quad . \label{eqmasterg} 
\end{equation}

In perturbation theory three types of
spin flips occur. As already stated above
only a single spin flips during the elementary process
$\bal\rightarrow \bbe$. From the study of the underlying 
CML ${\mathbf{T}}_{\eps , \delta}$ one can
infer the following properties of the transition probabilities 
$  w \leb \bbe \, | \, \bal \rib  $. 
\bei
\item Because of the nearest neighbour coupling in eq.~(\ref{eqcml}) and
the direct product property (\ref{eqprourbn4}) in the perturbative regime,
the transition probabilities $w(\bbe \, | \, \bal)$ depend on the three
neighbouring spins only
\bequ
w(\bbe \, | \, \bal) = w^{(3)} \left( \alpha_{i-1} \, \beta_{i} \,
\alpha_{i+1} \; | \; \alpha_{i-1} \, \alpha_{i} \, \alpha_{i+1} \right) \; 
\quad ,
\label{eqreuktuebw}
\end{equation}
where $ \alpha_i $ denotes the transition index.
Hence the spin interaction is local. 
\item According to eq.~(\ref{eqreuktuebw}) the 
transition probabilities $  w \leb \bbe \, | \, \bal \rib  $ 
do not depend on $N$. Therefore, they can be
determined for small systems, e.~g.~$N=3$.
\item Transitions of the three different types have probabilities $w_a$, 
$w_b$, and $w_c$, respectively, which
depend on the parameters $\eps, \, \delta$.
If $\delta $ is greater than the corresponding critical value in 
eq.~(\ref{eqcrival}), the respective
transition probability strictly vanishes. Lowering $\delta$ the transition
probability increases monotonously as can be
shown by a rather subtle argument which uses
the monotonous growth of the overlap set $OV_{\bal, \, \bbe}$.
\eei
The dynamics resulting from the master equation (\ref{eqmasterg}) is
almost trivial in parameter regions 1 and 2, since at most 
the spin flip of type (a) is possible. Regions 3 and 4 are more
interesting, because at least two different spin flips occur. 
Since we are mainly interested in large systems, we confine
ourselves to $N$ even in what  follows. 

In region 3 spin flips of type (a) and (b) are possible. On the 
coarse grained level the attractors  $I_{(\,  { +1 },  {-1},  
{ +1 },   {-1}, \dots ,  {-1}, { +1 },  {-1} \,)} $ and $ I_{(\,   {-1}, { +1
},   {-1},  { +1 }, \dots , { +1 },  {-1}, { +1 } \,)}$ are viewed as the two
ground states of the antiferromagnetic Ising model. Hence, the ergodic dynamics
of the CML ${\mathbf{T}}_{\eps , \delta}$ corresponds to an antiferromagnetic 
Ising model at zero temperature. In parameter region 4 all three spin flips
are possible. The (unique) stationary distribution of the master
equation can be calculated with the ansatz that the weight of each state 
$\bal$ solely depends on the number of defects. The result 
\begin{equation}
p^{stat}_{\bal} =  c    \left( \frac{w_c}{w_a} \right)^{\frac{1}{4} 
\sum_{i=1}^N  \alpha_i  \alpha_{i+1}  } 
= \frac{1}{Z}   \exp \left( \beta \,  J \,  \sum_{i=1}^N
\alpha_i  \alpha_{i+1} \right)
\label{eqgleigeisi}
\end{equation}
clearly can be cast into the form of a canonical distribution for
a nearest neighbour coupled Ising chain. 
Here, $c$ and $Z$ denote the normalisation
constants and for the temperature the relation
\begin{equation}
\beta \, J = \frac{1}{4}  \log \leb \frac{w_c}{w_a} \rib 
\label{eqbezisicml}
\end{equation}
follows. Taking $J$ with modulus one the temperature depends on
the ratio of the transition probabilities for 
generation and annihilation of two
defects. It is finite throughout region 4. Ferromagnetic coupling, $J=+1$,
is obtained for $(w_c / w_a) > 1$, whereas in the opposite case
$(w_c / w_a) < 1$ antiferromagnetic coupling, $J=-1$, follows.
Both cases are realized in the present 
parameter region, as can be seen in figure \ref{figbifu3}. Here,
the transition probabilities $w_a$ and $w_c$ were obtained 
numerically by analysing the transitions of a very long orbit of the CML
with $N=3$. Finally, it is easy to show that the stationary distribution 
(\ref{eqgleigeisi}) of eq.~(\ref{eqmasterg}) 
obeys detailed balance
\[
w( \bbe \, | \, \bal) \cdot p^{stat}_{\bal} =  
w( \bal \, | \, \bbe) \cdot p^{stat}_{\bbe}
\; ,
\]
although the underlying CML describes a non--equilibrium  process
on the microscopic level. 

\section{Transient dynamics of the CML} \label{sectrans}
As mentioned above the transient dynamics of the CML is most interesting in
the parameter region 3 for which the attractors $I_{(\,  { +1 },  {-1},  { +1
},   {-1}, \dots ,  {-1}, { +1 }, {-1} \,)} $ and $ I_{(\,   {-1}, { +1 },   
{-1},  { +1 }, \dots , { +1 },  {-1}, { +1 } \,)}$ have large basins of
attraction for large $N$. The {\em mean transient time} $\langle T \rangle $  
can be determined by averaging the time until an orbit reaches one of the two
attractors over many random initial conditions, i.e.\ initial conditions
distributed according to the Lebesgue measure. The numerical simulation
indicates a quadratic increase with the system size
\bequ
\langle T \rangle \propto N^2 \quad \quad (N \gg 1) \quad .
\label{eqasymtral}
\end{equation}
Such a law can be understood from the coarse grained
point of view. In parameter region 3 the spin flips of type
(a) and (b) are allowed which
cause the annihilation of two defects respectively 
the diffusion of one defect. Overall, the transient dynamics of the CML 
corresponds to a  relaxation process towards 
one of the two ground states. Since 
diffusion is important for the relaxation the time scale grows with
the second power of the length scale, i.e.\ the system size.

To be more definite and in order to 
derive eq.~(\ref{eqasymtral}) formally
we remind that the spin dynamics induced by the map lattice
constitutes a {\em kinetic Ising model} with local spin flips.
Models of these type have been introduced by 
R.~J.~Glauber in his celebrated article \cite{glau,kawas}.
So the coarse grained dynamics of the CML ${\mathbf{T}}_{\eps , \delta}$ belongs
to a well--studied class of models. For the case of zero temperature,
i.~e.\ $w_c=0$, an exact analytical solution is available if
an additional relation for the two remaining transition probabilities
is imposed
\bequ
w_a = 2 \, w_b \; \quad .
\label{eqwawcgl}
\end{equation} 
Such a condition holds only on a subset of parameter region 3, namely on a line
of $\eps, \, \delta$ values. One can show with the help of 
results on the temporal evolution of correlation functions that the mean
number of defects in the spin chain obeys
\bequ
\langle \# \, {\rm defects} \left( t \right) \rangle 
\sim \frac{1}{\sqrt{ 8 \, \pi \, w_b}} \; \frac{N}{\sqrt{t}}\, , \quad \quad 
\; \; t \gg 1 \;  \quad .
\end{equation} 
If one changes the system size from $N$ to $k \cdot N$ the
time $t$ has to be scaled by a factor $k^2$ in order to reach the same number
of defects. Since the mean transient time $\langle T \rangle $ 
determines the scale for the annihilation of all defects, the
scale argument implies relation (\ref{eqasymtral}).
  
When one relaxes condition (\ref{eqwawcgl}) between transition probabilities
the quadratic growth of the transient time with $N$ in eq.~(\ref{eqasymtral}) 
still holds as numerical simulations indicate. Such an observation is in
accordance with the theory of dynamical critical phenomena \cite{hohal}. The 
latter implies universal scaling laws for relaxation
phenomena at the critical point. At zero temperature the dynamics of a 
one dimensional kinetic Ising model is critical and the 
decay of defects is governed
by the dynamical critical exponent $z$
\bequ
\langle \# \, {\rm defects} \left( t \right) \rangle 
\sim  \frac{N}{t^{1/z}} \quad .
\end{equation}
$z$ equals two for kinetic Ising models where the order parameter 
is not conserved \cite{hohal,priv}. 
Consequently eq.~(\ref{eqasymtral}) holds for a large set
of parameter values in region 3.
\section{Summary}
We have introduced a coupled map lattice which was constructed in
analogy to the Miller Huse model. By a perturbation expansion
for weak coupling and in the vicinity of a symmetry breaking
bifurcation of the single site map nontrivial dynamical behaviour 
has been investigated. Our approach was based on analysing
geometric properties in phase space. Transitions between
certain cubes which are the building blocks of a coarse grained 
description have been computed. A global bifurcation of the dynamics 
occurs if a transition becomes allowed or forbidden by a change of 
the parameters. Four parameter regions with different ergodic behaviour
could be identified. As a surprising and counter intuitive
feature of our map lattice we mention that increasing the spatial coupling
inhibits transitions and stabilises single cubes as attractors. As a 
consequence the coupling acts somehow antiferromagnetic on a coarse grained
level.

Performing a coarse graining of the map lattice
the resulting symbol or spin dynamics becomes a kinetic Ising model. 
We have been able to identify parameter regions where our dynamical
system can be mapped to a finite temperature nearest neighbour
coupled Ising chain. Depending on the original parameters of the system
ferro-- or antiferromagnetic coupling can be realised, but
the ordered phase at zero temperature is always in the 
antiferromagnetic regime. 
The coarse grained viewpoint also sheds some light on the transient dynamics 
of the map lattice since the transients correspond to a
relaxation process in the kinetic Ising model. Therefore, the transient
behaviour of the CML is related to a non--equilibrium process of statistical
physics. 

Within our approach we have successfully linked the dynamics of a coupled
map lattice to properties of a kinetic Ising model on analytical grounds.
Of course our approach is not mathematically rigorous, but we have good
indication that the results are valid at least in the perturbative regime.
The comparison with numerical simulations 
shows that the leading order of perturbation 
theory is a good description for parameter values 
$\eps, |\delta| \lesssim 5 \cdot 10^{-2}$. 

For further studies the adaption of the method to coupled maps on a 
two dimensional lattice seems desirable since here also finite temperature
phase transitions are possible. This would constitute a
further step in the understanding of phase transitions in coupled map
lattices as exemplified by the Miller Huse model.
\appendix
\section{} \label{appreduct}
We want to derive eq.~(\ref{eqprourbn4}) for symbol sequences $\bal$, $\bbe$ 
with $\alpha_j=\beta_j$ for $j\neq i$ and $\alpha_i\neq \beta_i$.
First, we remind that the single site map $f_{\delta}$ is affine on the 
four intervals (cf.~figure \ref{figourmap}): 
\begin{equation}
K({-2}) := [-1, \, -a], \quad K({-1}) := [-a,\, 0] , \quad
K({1}) := [0,\, a], \quad  K({2}) := [a, \, 1]  \; .
\end{equation}
Consequently, the map $ {\mathbf{T}}_{{ \eps }, {\delta}} $ is affine on the
$4^N$ cuboids 
\bea
S_{\boldgamma } & := & K (\gamma_1) \times  K (\gamma_2) \dots  
\times K (\gamma_N) \nonumber \\
 \boldgamma & := &  (\gamma_1 , \gamma_2 \dots \gamma_N) \quad \; {\rm{with}}
\;\;  \gamma_i \in \{-2,-1,+1,+2 \} \quad .
\label{eqsubq}
\eea 
Each cube $I_{\bal}$ contains $2^N$ cuboids $S_{\boldgamma }$. The image of a
cuboid under $ {\mathbf{T}}_{{ \eps }, {\delta}} $ is a parallelepiped
\begin{equation}
P_{\bga} := {\mathbf{T}}_{{ \eps }, {\delta}} \leb S_{\boldgamma }\rib \; 
\end{equation}
which is a weakly deformed cube $I_{\bal}$ for $S_{\bga} \subset I_{\bal}$, 
because ${\mathbf{T}}_{{ \eps } = {\delta} = 0} \leb
S_{\bga} \rib = I_{\bal}$ holds and we are in the perturbative regime $\eps, \,
|\delta | \ll 1$. The distances between the corners of $P_{\bga}$ and of
$I_{\bal}$ are of the order $\olo$. 

The overlap set $ OV_{\bal, \, \bbe} $ as defined in eq.~(\ref{overlapset}) 
then reads 
\bequ
OV_{\bal, \, \bbe} = \bigcup_{\{\bga \; | \;S_{\boldgamma} 
\subset I_{\boldalpha}\} } (\, P_{\bga} \cap I_{\bbe} \,) \; .
\label{eq"ubasn5}
\end{equation}
The intersection of a parallelepiped $ P_{\bga} $ with $I_{\bbe}$ can be
written as
\bequ
P_{\bga} \cap I_{\bbe} = \Big \{ {\mathbf{T}}_{{\eps}, { \delta}}({\mathbf{x}})
\; \Big | \; {\mathbf{x}} \in S_{\bga} \, \wedge \, \prod_{j=1}^N \; 
\theta \left( \beta_j \,  \left[ {\mathbf{T}}_{{\eps}, { \delta}}({\mathbf{x}})
\right]_{j}  \right) = 1  \Big \} \;  \quad ,
\label{eqdarsnn4}
\end{equation}
where $\theta (x) $ denotes the Heaviside function. 

Since $\beta_j = \alpha_j$ for $j \neq i $ we have
\begin{equation}
\theta \left( \beta_j \,  \left[ {\mathbf{T}}_{{\eps}, { \delta}}(\mathbf{x})
\right]_{j}  \right) = 1 
\end{equation}
provided $x_j$ has at least a distance of order $\olo$ from
the endpoints of the interval $K \leb \gamma_j \rib$.
Therefore the set (\ref{eqdarsnn4}) can be approximated by
\bequ
P_{\bga} \cap I_{\bbe} = \left\{ {\mathbf{T}}_{{\eps}, {
\delta}}({\mathbf{x}}) \; \left| \; {\mathbf{x}} \in S_{\bga} \, \wedge \, 
\theta \left( \beta_i \,  \left[ {\mathbf{T}}_{{\eps}, { \delta}}({\mathbf{x}})
\right]_{i} \right) = 1 \right. \right\} \; 
\label{eqapprvol4a}
\end{equation}
in leading order of perturbation theory. The remaining
Heaviside function in eq.~(\ref{eqapprvol4a}) only depends on the
coordinates $x_{i-1}$, $x_i$ and $x_{i+1}$ because of the local coupling of
the CML $ {\mathbf{T}}_{{\eps}, {\delta}} $. Therefore and since $P_{\bga}$
is a weakly deformed cube $I_{\bal}$, we obtain in the same order
of approximation
\bea
P_{\bga} \cap I_{\bbe} 
&= & \left\{ 
{\mathbf{T}}^{(3)}_{{\eps}, {\delta}}({\mathbf{x}}^{(3)}) 
\; \Big | \; {\mathbf{x}}^{(3)} \in S^{(3)}_{ \gamma_{i-1} \,
\gamma_{i} \, \gamma_{i+1}} \, \wedge \; \theta \left( \beta_i \,  \left[
{\mathbf{T}}^{(3)}_{{\eps}, { \delta}}({\mathbf{x}}^{(3)}) 
\right]_{i} \right) = 1
\right\} 
\times  I^{(N-3)}_{\alpha_1 \,\alpha_2 \dots \alpha_{i-2} \, 
\alpha_{i+2} \dots \alpha_{N}} \nonumber \\ 
&=& \leb P^{(3)}_{ \gamma_{i-1} \, \gamma_{i} \, \gamma_{i+1}}  
\cap I^{(3)}_{\beta_{i-1} \, \beta_i \, \beta_{i+1}} \rib
\times I^{(N-3)}_{\alpha_1 \,\alpha_2 \dots \alpha_{i-2} \, 
\alpha_{i+2} \dots \alpha_{N}} \; .
\label{eqprodsnn}
\eea
Here the superscript indicates that the quantities are determined by
a map lattice of size $N=3$ with ${\mathbf{x}}^{(3)}=
(x_{i-1} , \, x_i , \, x_{i+1})$ .
The $(N-3)$ dimensional cube $I^{(N-3)}_{\alpha_1 \,\alpha_2 
\dots \alpha_{i-2} \, \alpha_{i+2} \dots \alpha_{N}}$ takes
the remaining coordinates $x_j$ with $j\not 
\in \{i-1, \, i, \, i+1 \}$ into account. 

If one approximates the map $ {\mathbf{T}}_{{\eps}, {\delta}} $ by the
simplified map 
\bea
\left[ \widetilde{ \mathbf{T}}_{ \eps , \delta } ({\mathbf{x}}
) \right]_i & = & 
 (1- \eps )  f_{\delta}(x_i)+ {\displaystyle \frac{\eps }{2}} 
( f_{\delta}(x_{i-1}) +
f_{\delta}(x_{i+1}) )  \nonumber \\
\left[ \widetilde{ \mathbf{T}}_{ \eps , \delta } ({\mathbf{x}}) \right]_j 
& = & 
f_{\delta}(x_j)\, , \;\;  \quad \forall j \neq i \; ,
\label{eqvereinfn4}
\eea
one arrives at the result (\ref{eqprodsnn}) at once. One can use
the simplified map $\widetilde{ \mathbf{T}}_{ \eps , \delta }$ for the
calculation of the pre--image sets ${\mathbf{T}}_{{ \eps }, {\delta}}^{-k}
(OV_{\bal, \, \bbe})$ in leading order perturbation theory. Since $\widetilde{
\mathbf{T}}_{ \eps , \delta } $ couples only the coordinates 
$x_{i-1}$, $x_i$ and $x_{i+1}$, in this approximation 
also the pre--image sets ${\mathbf{T}}_{{ \eps}, 
{\delta}}^{-k} (OV_{\bal, \, \bbe})$ have the structure of a direct 
product in eq.~(\ref{eqprourbn4}).
Finally, it can be easily shown that the 
blind volume $ B_{\bal}$ of the cube $I_{\bal}$ can also be
approximated by a direct product of the form as in eq.~(\ref{eqprourbn4}). 

\section{} \label{appdeltac}
In order to illustrate the main steps for the calculation of the
critical values $\delta_{crit.}(\eps)$ we focus on the case
$N=2$ and the transition $I_{++} \rightarrow I_{-+}$. 
Generalisations to $N>2$ and different transitions are almost
obvious, but require some tedious though elementary computations \cite{diss}.

For calculating the overlap set $OV_{-+, ++}$ we 
introduce the following shorthand notation for the indices of the rectangles 
in eq.~(\ref{eqsubq}) 
\begin{equation}
S_1:=S_{-2\, 1}, \; S_2:=S_{-2\, 2}, \; S_3:=S_{-1\, 1}, \; S_4=S_{-1\, 2}
\quad .
\label{eqshorth}
\end{equation}
With the parallelogram $P_i :={\mathbf{T}}_{{\eps}, { \delta}}(S_i)$ 
the overlap set (\ref{eq"ubasn5}) reads
\bequ
OV_{-+, ++} = \bigcup_{i=1}^{4} (\, P_{i} \cap I_{++} \,) \; .
\label{eq"ubasn4}
\end{equation}
Within first order $P_1 \cap I_{++}$ and $P_2 \cap I_{++}$ as well 
as $P_3 \cap I_{++} $ and $P_4 \cap I_{++} $ are
equal to each other. $P_3$ and the intersection
$P_3 \cap I_{++}$ are shown in figure \ref{figpara3}. The area of the
latter triangular set is $\eps / 2$ in first order. The intersection
$P_1 \cap I_{++}$ is obtained by shifting the just mentioned triangle by an
amount $-\delta$. 
\bef
\begin{center}
\mbox{\epsfig{figure=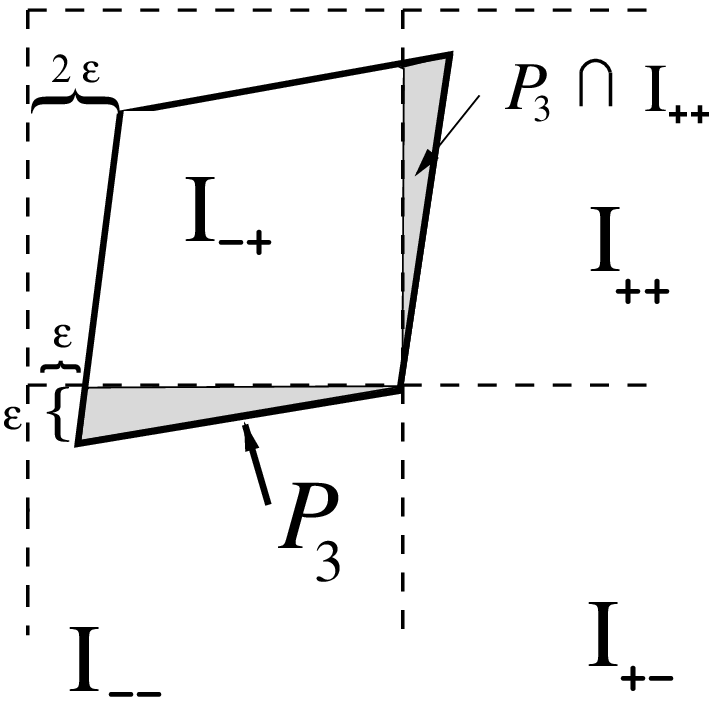,width=40mm}}
\end{center}
\caption{The parallelogram $P_{3} = {\mathbf{T}}_{{\eps}, { \delta}} (S_3)$ 
and its intersection with the square $I_{++}$.}
\label{figpara3}
\eef

The pre--image set of the overlap, ${\mathbf{T}}_{{\eps}, 
{ \delta}}^{-1} \leb OV_{-+, ++} \rib $, is displayed in figure
\ref{figpreimage} where we restrict the parameter range to
$-2 \, \eps < \delta < 0$ for simplicity. For calculating pre--images of
higher order the so called ''blind area'' $B_{-+}$ 
comes into play, i.~e.\
the set of points in $I_{-+}$
which do not have pre--images with respect to the map
${\mathbf{T}}_{{\eps}, { \delta}}$. 
The components of ${\mathbf{T}}_{{\eps}, { \delta}}^{-1}
\leb OV_{-+, ++} \rib $ in $S_1$ and $S_2$ are contained in
the subset $T$ of the blind area $B_{-+}$ (cf.~figure \ref{figpreimage}).
In first order the width of the set $T$ is given by the expression 
\bequ
b_{T} \leb x_2 \rib = \eps + \eps \, x_2 \, , \quad x_2 \in  [0,1]  \; . 
\label{eqbrfv}
\end{equation}
\bef
\begin{center}
\mbox{\epsfig{figure=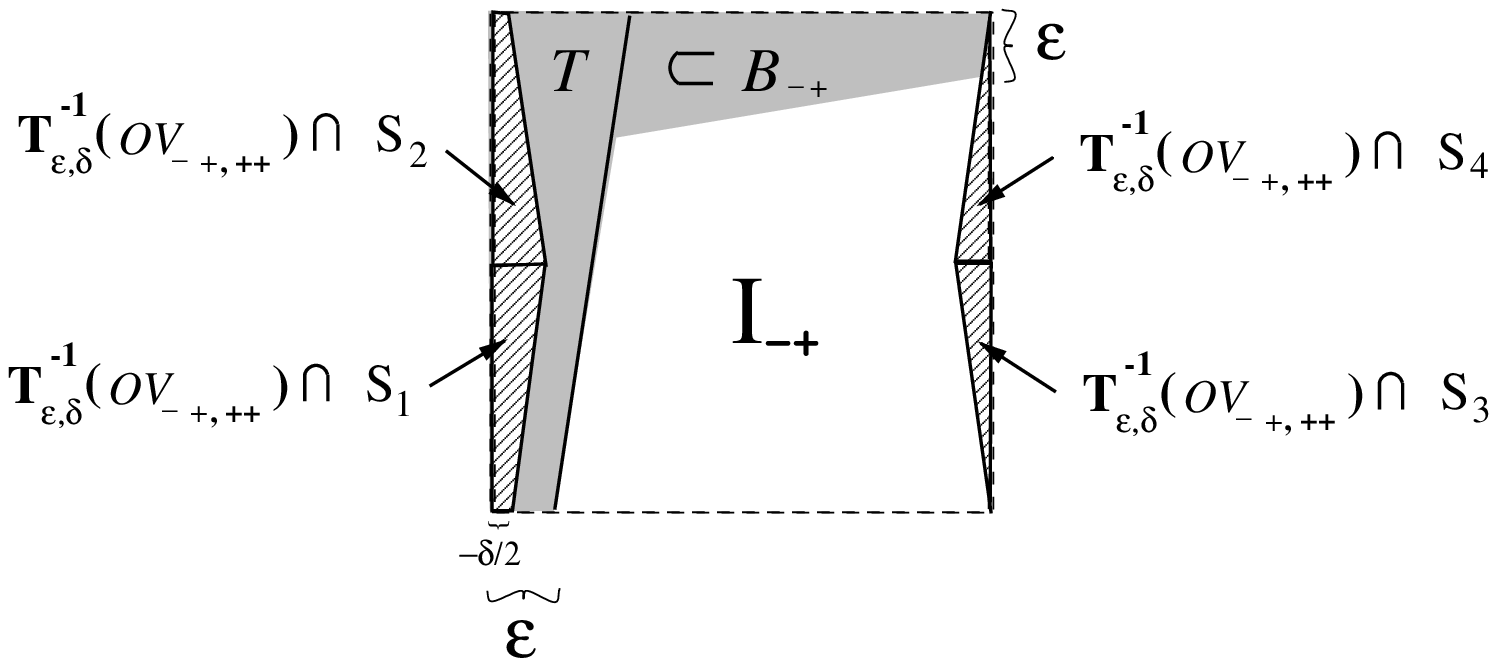,width=100mm }}
\end{center}
\caption{The pre--image set  ${\mathbf{T}}_{{\eps}, { \delta}}^{-1} \leb
OV_{-+, ++} \rib $ and its four components in the rectangles $S_i$
($-2 \, \eps < \delta < 0$). Additionally the blind area $B_{-+}$ and 
its subset $T$ (cf.\ eq.~(\ref{eqbrfv})) are displayed in gray.
\label{figpreimage}}
\eef

For convenience in calculating pre--images of higher order we first concentrate
on the right rectangles $ S_3 $ and $ S_4 $. 
Defining the generations of order $k$ by 
\bea 
G^{(1)} & := & {\mathbf{T}}^{-1}_{{ \eps },
{\delta}}(OV_{-+,++}) \cap ( \, S_{3} \cup S_{4} \,) \; ,
 \nonumber \\
G^{(k)} & := & \left\{ {\mathbf{x}} \in  \left( \, S_{3} \cup S_{4} \,
 \right)\;  | \; {\mathbf{T}}_{{ \eps }, {\delta}}(
{\mathbf{x}}) \in G^{(k-1)}  \right\}\; , \quad \; k=2,3, \dots
\label{eqdefgk}
\eea
figure \ref{figdreig} reveals a beautiful recursive structure of these sets.
In order to describe this structure analytically we remark that up to
first order it is sufficient to compute pre--images with respect
to a simplified map (cf.\ eq.~(\ref{eqvereinfn4}))
\bea 
\left[ \widetilde{{\mathbf{T}}}_{ \eps,\delta } 
({\mathbf{x}}) \right]_1 & = & 2 {x}_1 
+ \eps f_{0}(x_2) \nonumber \\
\left[ \widetilde{{\mathbf{T}}}_{ \eps,\delta } 
({\mathbf{x}}) \right]_2 & = & f_{0}({x}_2)  
\quad .
\label{eqverei3}
\eea
Then the following properties of the generations $G^{(k)}$, 
which are inherent in figure \ref{figdreig}, are easily obtained
\bei
\item The first generation $ G^{(1)} $ consists of two triangles 
with vertices $(0,0)$, $(0, 1/2)$, $(-\eps/2, 1/2)$
respectively $(0,1)$, $(0, 1/2)$, $(-\eps/2, 1/2)$.
\item A generation $ G^{(k)} $ encompasses $2^k$ triangles each of them having
the same area. The area shrinks by a factor 4, if one passes from $G^{(k-1)}$
to $ G^{(k)} $.
\item Two neighbouring triangles of the same generation share a corner or a 
side with length of order $\eps$.
\item The union $\Sigma_{G}^{(k)} := \cup_{n=1}^{k} \, G^{(n)} $ 
is a simply connected set.
\eei
To determine the boundary of $\Sigma_{G}^{(k)}$ we consider its
height function
\bequ
R^{(k)} (x_2) := \inf \left\{ \, x_1 \; | \; ( x_1, \, x_2 ) \in
\Sigma_{{G}}^{(k)} \; \right\} \; .
\label{eqdefrkge}
\end{equation}
Since $R^{(k+1)}$ is mapped on  $R^{(k)} $ by the simplified map 
$\widetilde{{\mathbf{T}}}_{{ \eps }} $, we get the representation
\begin{equation}
R^{(k)}(x_2)  =  - \eps \; \sum_{i=1}^{k} \frac{ f_0^i(x_2)}{2^i} 
\, , \quad \quad x_2 \in [ 0 , 1 ] \quad .
\label{eqrkallg}
\end{equation}
For $k$ odd these curves admit $ 2^{(k-1)/2} $ absolute extrema
at
\bequ
x_{min} \in
\left\{ \frac{1}{2} \, \left( 1 + \sum_{j=1}^{(k-1)/2} 
\frac{i_j}{4^j} \right) \; \bigg| \quad  i_j \in \{-1,+1 \} , \; j=1,2,
\dots(k-1)/2 \right\}   
\label{eqx2max}
\end{equation}
with height
\bequ
R^{(k)} (x_{min})  = - 
\frac{\eps}{2}\, \sum_{i=0}^{(k-1)/2} \, \frac{1}{4^i} \; .
\label{eqhoeend1}
\end{equation}
In the limit $k \to \infty $ the set 
$\Sigma_G^{\infty}$ has a fractal
boundary, since its construction is
analogous to the famous Koch's curve \cite{mandel}. The thickness of the set
$\Sigma_G^{\infty}$ follows easily from eq.~(\ref{eqhoeend1}) 
\begin{equation}
h \left( \Sigma_{{G}}^{\infty} \right) := {\rm sup} 
\left\{ \, |x_1| \; | \; 
{\mathbf{x}} \in  \Sigma_{{G}}^{\infty} \, \right\} = \frac{2 \, \eps}{3} \; .
\label{eqheigin}
\end{equation}
\bef
\begin{center}
\mbox{\epsfig{figure=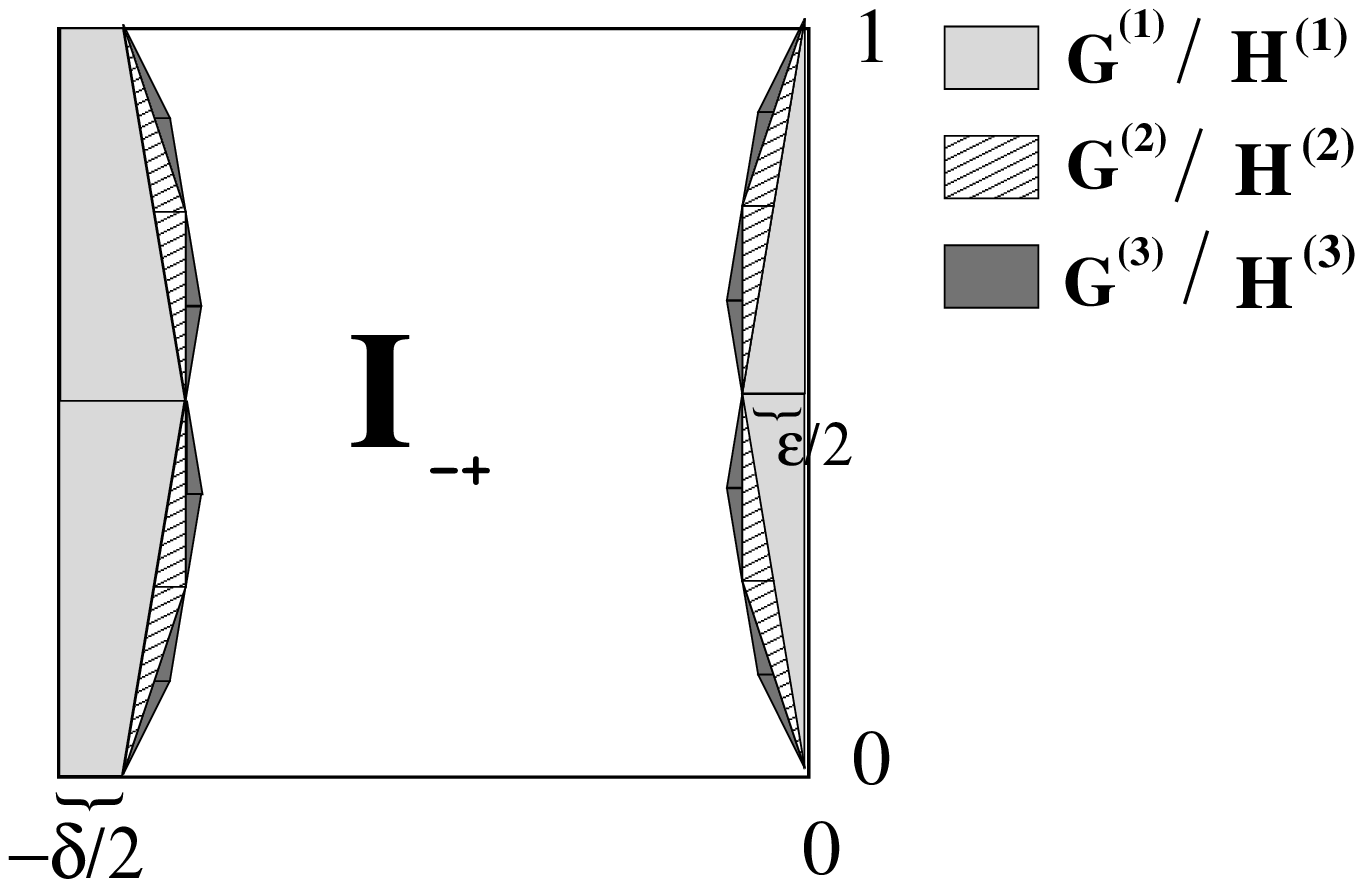,width=80mm} }
\end{center}
\caption{The first three generations $G^{(k)}$ and $H^{(k)}$ 
near the right resp.~left edge of $I_{-+}$ \label{figdreig}.}
\eef

A generation $ {G}^{(k)} $ has not only pre--images $ {G}^{(k+1)} $ in the
right rectangles $S_3$ and $S_4$, but there are also
pre--images in the left rectangles $S_1$ and $S_2$ 
(cf.~figure \ref{figdreig})
\bea
H^{(1)} & :=  & {\mathbf{T}}^{-1}_{{ \eps }, {\delta}}(OV_{-+,++})\, \cap \,
 (\,S_{1} \cup S_{2} \,) \, , \nonumber \\ 
H^{(k)} & := & \left\{ {\mathbf{x}} \in  \left( \, S_{1} \cup S_{2} \,
 \right)\;  | \; {\mathbf{T}}_{{ \eps }, {\delta}}(
{\mathbf{x}}) \in G^{(k-1)}  \right\}\, , \quad \; k=2,3, \dots \quad .
\label{eqdefh}
\eea
To reveal the relation between $G^{(k)}$ and $H^{(k)}$ analytically
we just note that for a point ${\mathbf y } \in  G^{(k-1)}$ 
eqs.~(\ref{eqdefgk}) and (\ref{eqdefh}) imply that
\[
{\mathbf{T}}_{{ \eps }, {\delta}} ({\mathbf{x}}) = 
{\mathbf{T}}_{{ \eps }, {\delta}} ({\mathbf{x'}}) = {\mathbf{y}} \in G^{(k-1)}
\, , \quad \; {\mathbf{x}} \in G^{(k)} , \; {\mathbf{x'}} \in H^{(k)} \; .
\]
Then to first order
\begin{equation}
x_1' + 1  =  - \delta /2 - {x}_1 , \quad
x_2'  =  x_2
\label{eqbedab}
\end{equation}
follows. Hence, the set $H^{(k)}$ is obtained from $G^{(k)}$ by a reflection
and an additional offset of $-\delta/2$ (cf.\ figure \ref{figdreig}). 
The same property follows of course for the limits $\Sigma_H^{\infty}$ 
and $\Sigma_G^{\infty}$. 

If
\bequ
\Sigma_{H}^{\infty} \subset  B_{-+} \; ,
\label{eqbeddc}
\end{equation}
holds, no further pre--images of the overlap set 
$OV_{-+, ++}$ appear, and the union $\Sigma_G^\infty\cup
\Sigma_H^\infty$ encompasses all pre--images.
Consequently, the transition $I_{-+} \rightarrow I_{++}$
is not possible, because all pre--images of the overlap 
set are located near the
edge of $I_{-+}$. Therefore,
condition (\ref{eqbeddc}) gives the clue for the determination of
$\delta_{crit.}(\eps)$. 

According to eqs.~(\ref{eqbedab}) and 
(\ref{eqheigin}) the thickness of $\Sigma_{H}^{\infty}$ reads
\bequ
h \left( \Sigma_{H}^{\infty} \right) := {\rm sup} \{ \, 1 + x_1 \; | \; 
{\mathbf{x}} \in  \Sigma_{H}^{\infty} \, \} = - \frac{\delta}{2}  +
\frac{2 \, \eps}{3} \; .
\label{eqheishi}
\end{equation}
At the critical value $\delta_{crit.}(\eps)$ one peak at the boundary
with maximal height
collides with the right border of the set $T \subset B_{-+}$
(cf.\ figure \ref{figdurch}). Since the boundary of the blind area
has according to eq.~(\ref{eqbrfv}) a finite slope, the peak
with the smallest $x_2$ coordinate crosses the right boundary of $T$ at 
first\footnote{This
can be shown rigorously with the inequality
\[
\sup \left\{ x_1 \; | \; (x_1, \, x_2 ) \in \Sigma_{H}^{\infty} \, 
\right\} \; \leq \; 
\sup \left\{ x_1 \; | \; (x_1, \, 1/3) \in \Sigma_{H}^{\infty} \, \right\} 
\, + \eps \, \left( x_2 - \frac{1}{3} \right) \; \quad .
\] 
}. According to eq.~(\ref{eqx2max}) this peak is located at $x_2 = 1/3$.
Then eqs.~(\ref{eqheishi}) and (\ref{eqbrfv}) yield
\bequ
\frac{2}{3} \, \eps - \frac{\delta_{crit.}(\eps)}{2} =
h \left( \Sigma_{H}^{\infty} \right)  = 
b_{T} \left(x_2 = 1/3 \right) = \frac{4}{3}  \eps 
\end{equation} 
and  consequently we arrive at
\bequ
\delta_{crit.}(\eps) = -  \frac{4}{3} \eps \; .
\end{equation}
\bef
\begin{center}
\mbox{\epsfig{figure=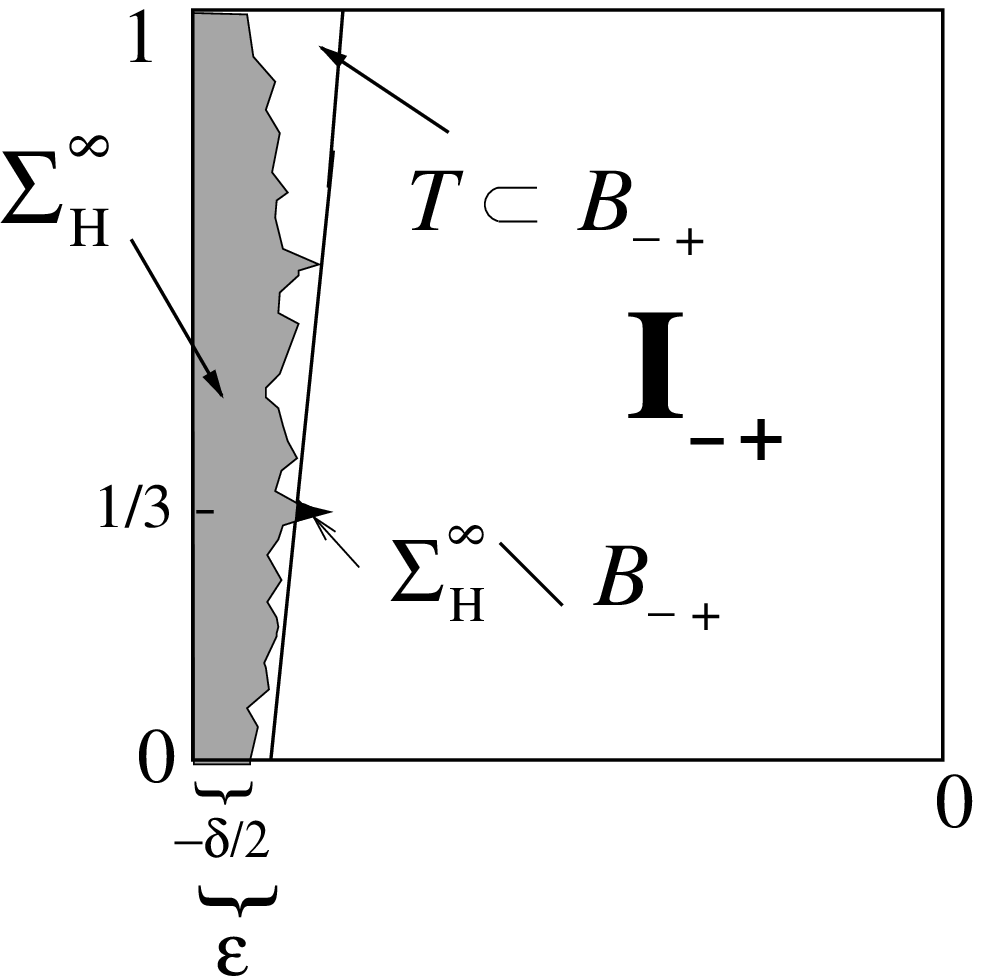,width=50mm}}
\end{center}
\caption{Diagrammatic view of the set $\Sigma_{H}^{\infty}$
and the subset $T\subset B_{-+}$ of the blind volume for 
$\delta < \delta_{crit.}(\eps )$. \label{figdurch} } 
\eef

In order to show that the 
transition $I_{-+} \rightarrow I_{++}$ is possible
for $\delta < \delta_{crit.}(\eps)$ the two conditions mentioned at the end
of section \ref{secmodel} have to be checked. Since $\Sigma_{H}^{\infty} 
\setminus B_{-+}$ is non--empty for $\delta<\delta_{crit.}(\eps)$, there
exists an open neighbourhood of $(x_1,x_2)=(-1,1/3)$ which is contained
within the pre--image set ${\mathbf{T}}_{{\eps}, {\delta}}^{-k_0} 
\leb OV_{-+, ++}\rib$ for a particular value $k_0$. Next
pre--images are located near $(-1/2,1/6)$ and $(-1/2,5/6)$ and hence
enter the inner part of the square $I_{-+}$. For higher generations again
four pre--images exist. Therefore it is plausible --
and more intricate considerations of \cite{diss} confirm it  --
that the set $\cup_{k=0}^{\infty} \, {\mathbf{T}}_{{\eps}, 
{\delta}}^{-k} \leb  OV_{-+, ++} \rib $ has a substantial Lebesgue measure.
For the second transition criterion we have to check whether the points that
are mapped into the overlap $OV_{-+,++}$ can migrate into the inner part of
$I_{++}$ under further iteration. A point $ {\mathbf x} \in OV_{-+,++}$ has a
positive $x_1$ coordinate of order ${\cal O}(\eps )$.
If one considers the evolution of the $x_1$ coordinate 
(cf.\ eq.~(\ref{eqverei3})), its value grows for most points 
$ {\mathbf x} \in OV_{-+,++}$
under further iteration, until it reaches a value of order one.
Therefore, the iterates reach the inner part of $I_{++}$ after a finite number
of steps. 

In conclusion, the transition $I_{-+} \rightarrow I_{++}$
becomes possible for $\delta < \delta_{crit.} \leb \eps \rib$.
Computation for other transitions or a CML with $N=3$ follows the same lines. 
We stress that the main steps consist in the calculation
of images and pre--images of overlap sets. The location of the
pre--image sets relative to the blind volume determines  
whether all pre--images of the overlap set 
are located near the edge of the cube only.
Consequently, the existence of the blind volume 
influences the numerical value of $\delta_{crit.}(\eps)$ significantly.  

\section{} \label{apphighord} 
{\new 
In this appendix we would like to show that for $N=2$ the cubes $I_{-+}$ and
$I_{+-}$ contain attractors in the strict mathematical sense, if $\delta >
- 4 \, \eps / 3$. Because of symmetry we can
concentrate on the cube $I_{-+}$. In appendix \ref{appdeltac} we have shown
that the (dominant) transitions $I_{-+} \rightarrow I_{++}$ and $I_{-+}
\rightarrow I_{--}$ are forbidden, as long as $\delta > \delta_{crit.} \approx
- 4 \, \eps / 3$. What remains to be done is that also the off diagonal 
transition $I_{-+}\rightarrow I_{+-}$ does not appear.

If $\delta < 0$ there exists a non--empty overlap set. Considering the
four rectangles in $I_{-+}$ on which the CML 
$ {\mathbf{T}}_{{\eps}, {\delta}} $ is linear (cf.\ eq.~(\ref{eqsubq})), 
only  the image of the rectangle $S_{-2 \, 2} := [-1, \, -a] \times 
[a, \, 1]$ intersects the square $I_{+-}$ for $\delta < 0$. 
\bef
\begin{center}
\mbox{\epsfig{figure=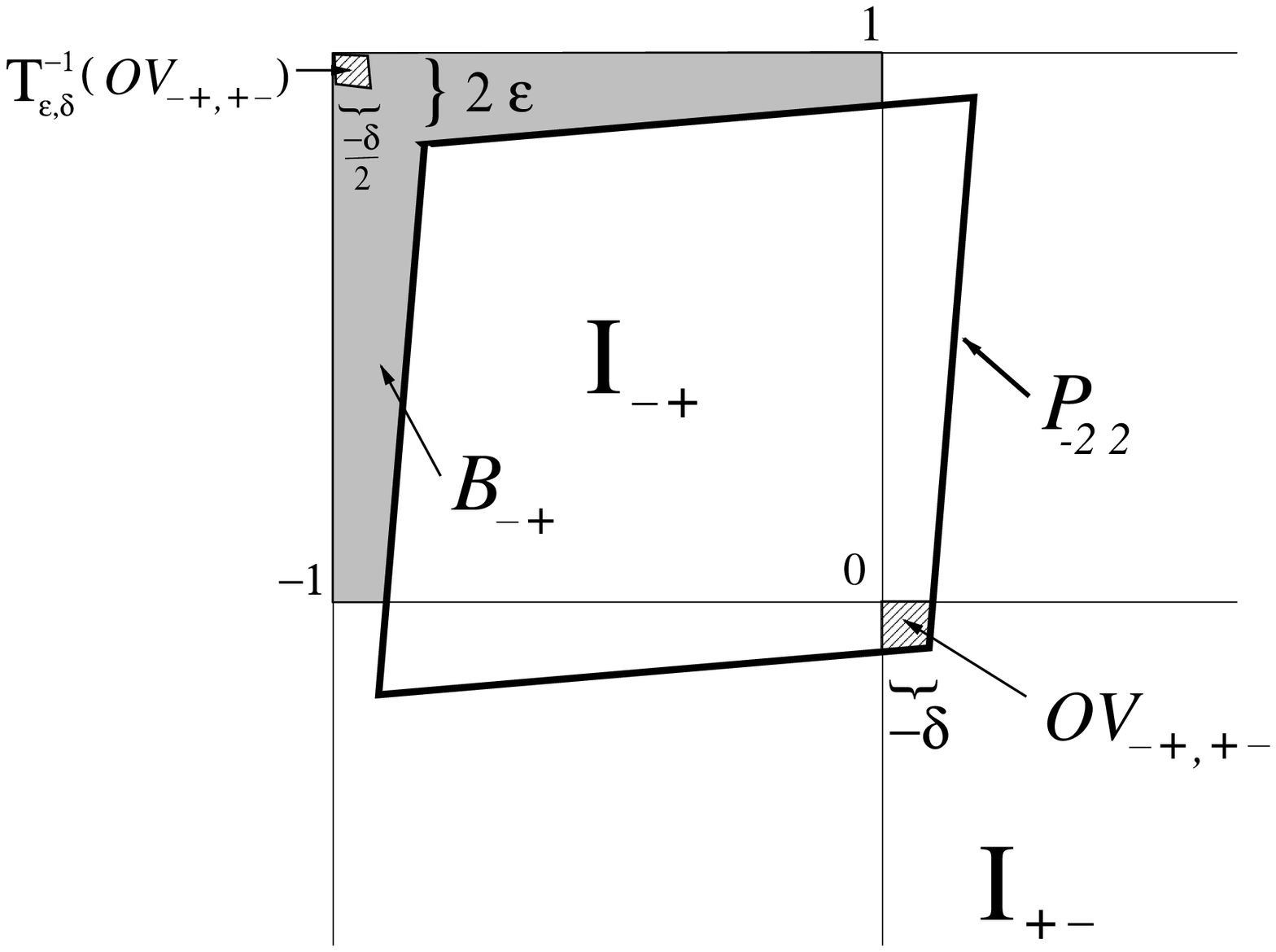,width=80mm}}
\end{center}
\caption{The overlap set $OV_{-+,+-}$ and its pre--image ${\mathbf{T}}_{{\eps},
{\delta}}^{-1}(OV_{-+,+-})$.\label{figpara5}}
\eef
Figure \ref{figpara5} displays this situation where the parallelogram 
$ P_{-2 \, 2} = {\mathbf{T}}_{{\eps}, {\delta}} (S_{-2 \, 2})$
and the overlap set
\begin{equation} \label{ovset}
OV_{-+,+-} = P_{-2 \, 2} \cap I_{+-}
\end{equation}
are shown. 
The overlap set has extension $\approx - \delta$ in the 
directions of both coordinate axes and hence an area $\approx \delta^2$.
Note that this area is factor of the order $\olo $ smaller than those of the
overlap sets $OV_{-+,++}$ and $OV_{-+,--}$ which belong to perturbatively
dominant transitions. 

The overlap set (\ref{ovset}) alone does not ensure for a transition
$I_{-+}\rightarrow I_{+-}$. In fact we show that phase space trajectories do
not reach this set, so that the transition does not appear. Since we are
considering a map with a finite coupling $\eps>0$, 
trajectories do not fill the whole phase space $[-1,1]^2$ but only
the subset ${\mathbf{T}}_{{\eps}, {\delta}}([-1,1]^2) \subset [-1,1]^2$.
In particular points close to the upper left corner of $I_{-+}$ are not visited.
This forbidden domain, previously called the blind volume, constitutes 
the reason why the off diagonal transition does not appear even beyond
the perturbation theory.

To put the argument on a formal level
we construct the pre--image sets of the overlap set within the square
$I_{-+}$. The first generation set, ${\mathbf{T}}_{{\eps},{\delta}}^{-1}
(OV_{-+,+-})$ is located near the corner $(-1, \, 1)$ of $I_{-+}$ and 
has sides 
of length $\approx - \delta / 2$ (cf.~figure \ref{figpara5}). As also shown in
this figure, the blind area $B_{-+}$ of the square $I_{-+}$ 
is also located there and contains a square with side length $2 \, \eps$. 
Therefore, the pre--image set ${\mathbf{T}}_{{\eps},{\delta}}^{-1} 
(OV_{-+,+-})$ is contained in the blind area, as long as $\delta > - 4 \,
\eps$. Hence in this parameter regime the overlap set has no pre--image sets 
${\mathbf{T}}_{{\eps},{\delta}}^{-k} (OV_{-+,+-})$ with $k \geq 2$, because
points belonging to the blind area have no pre--images themselves. In
particular, the pre--images of the overlap set do not intersect the inner
part of the square $I_{-+}$, so that one criterion for the transition 
$I_{-+}\rightarrow I_{+-}$ is not obeyed and the transition is impossible 
for $\delta > - 4 \, \eps$. 

Summarising, none of the transitions 
$I_{-+}\rightarrow I_{\bal}$ with $\bal \in \{ --,
\; ++, \; +-  \} $ is possible for $\delta > \delta_{crit.} \approx - 4 \, 
\eps / 3$. Therefore, in this parameter region attractors in the strict sense
reside within the squares $I_{-+}$ and $I_{+-}$.
}
\bibliographystyle{prsty}

\end{document}